\documentclass[fleqn,twoside]{article}
\usepackage{amsmath}  
\usepackage{amssymb} 
\usepackage{espcrc2}

\usepackage{graphicx}
\usepackage[figuresright]{rotating}

\newcommand{\be}{\begin{equation}}
\newcommand{\ee}{\end{equation}}
\newcommand{\bes}{\begin{equation*}}
\newcommand{\ees}{\end{equation*}}

\title{Deconfinement transition in 2+1-dimensional SU(4) lattice gauge theory}

\author{Philippe de Forcrand\address[ETHZ]{Institute for Theoretical Physics, ETH Z\"{u}rich, CH-8093 Z\"{u}rich, Switzerland}\address[CERN]{CERN, Theory Division, CH-1211 Geneva 23, Switzerland}\thanks{Talk presented by Ph. de Forcrand} and
        Oliver Jahn\addressmark[ETHZ]}

\begin{document}

\begin{abstract}

A missing piece is added to the Svetitsky-Yaffe conjecture. The spin model in the same universality
class as the $(2+1)d$ SU(4) theory, the $2d$ Ashkin-Teller model, has a line of
continuously varying critical exponents. 
The exponents measured in the gauge theory correspond best to the Potts point on the Ashkin-Teller line.

\vspace{-0.5pc}
\end{abstract}

\maketitle

\section{FRAMEWORK}

The $SU(4)$ Yang-Mills theory in $(2+1)$ dimensions has a special status:
its critical exponents at the finite temperature deconfinement transition
are not specified by the Svetitsky-Yaffe (S-Y) conjecture \cite{SY}. By measuring
them, we can learn about the couplings of the equivalent spin
model, i.e. of the effective Polyakov loop model.

The S-Y conjecture says that, {\em if} the deconfinement transition of the 
gauge theory is second-order, then it is in the same universality class as 
the spin model having the symmetry-breaking pattern of the Polyakov loop.

Assume that the $(2+1)d$ $SU(4)$ deconfinement transition is 
second-order -- we will investigate this issue numerically next. Then the
equivalent $2d$ spin model has $Z(4)$ spins, with orientation
$\{0,\pm\frac{\pi}{2},\pi\}$ corresponding to the 4 Polyakov loop sectors
and the 4 perturbative vacua of the gauge theory. A nearest-neighbour
Hamiltonian, which gives the same critical properties as the $SU(4)$ theory
according to S-Y, will have 3 possible energy levels for
each link: $E_0$ for parallel spins $\uparrow \uparrow$, 
$E_1$ for perpendicular spins $\uparrow \rightarrow$, and
$E_2$ for anti-parallel spins $\uparrow \downarrow$. Remarkably, the critical exponents of this
spin system are known to vary {\em continuously} with the ratio $\rho \equiv
\frac{E_2 - E_1}{E_1 - E_0}$ and the corresponding couplings of the 
Hamiltonian. Therefore, the S-Y conjecture does not tell us what the $SU(4)$
critical exponents are. Measuring them fixes the couplings of the
effective Hamiltonian.

A $Z(4)$ spin model with 3 energy levels per link is called a symmetric
Ashkin-Teller model \cite{AT}. It is the symmetric case $J=J'$ of the 
Ashkin-Teller model, which describes 2 coupled Ising systems 
$\{\sigma_i,\tau_i\}$ on a square lattice, with Hamiltonian:
\be
H = -J \sum_{<ij>} \sigma_i \sigma_j -J' \sum_{<ij>} \tau_i \tau_j
-K \sum_{<ij>} \sigma_i \tau_i \sigma_j \tau_j
\label{H_AT}
\ee
When $J=J'$, it can be rewritten
\be
H = -2J \sum_{<ij>} \cos(\theta_i - \theta_j) -K  \sum_{<ij>} \cos(2(\theta_i - \theta_j))
\ee
where now $\theta_i=\{0,\pm\frac{\pi}{2},\pi\}$ represents the orientation
of an $SU(4)$ Polyakov loop. The ratio of excitation energies mentioned
above, $\rho \equiv \frac{E_2 - E_1}{E_1 - E_0}$, is then equal to $\frac{J-K}{J+K}$.
Two special cases arise: \\
$\bullet$ the 4-state Potts case: $K=J \Rightarrow \rho=0$. No additional
energy is needed to make an orthogonal spin pair anti-parallel. \\
$\bullet$ the Ising$^2$ case: $K=0 \Rightarrow \rho=1$. It takes twice as
much energy to flip one spin in an aligned pair, as to make it orthogonal.
The name comes from the equivalent description as two decoupled Ising systems
in Eq.(\ref{H_AT}). \\
On physical grounds, one expects $0 \leq \rho \leq 1$, so that the two cases
above are limiting cases. As $\rho$ varies in the interval, the critical 
exponents also do, between the following limits:
\begin{center}
\begin{tabular}{|c|c c c c c c|}
\hline 
          & $\gamma/\nu$ & $\gamma'/\nu$ & $\beta/\nu$ & $\beta'/\nu$ & $\alpha/\nu$ & $\nu$ \\
\hline 
Potts     &       7/4    &        7/4    &      1/8    &       1/8    &        1     &   2/3 \\
Ising$^2$ &       7/4    &        3/2    &      1/8    &       1/4    &        0     &    1  \\
\hline 
\end{tabular}
\end{center}
\noindent
where
$\gamma'$ and $\beta'$ are susceptibility and magnetization exponents for spins $e^{2 i\theta}$.
Since $\gamma/\nu$ and $\beta/\nu$ are independent of $\rho$, these known exponents provide
crosschecks on our systematic errors. This is particularly valuable here,
because of large, logarithmic finite-size corrections known to make the 
numerical extraction of Ashkin-Teller exponents quite challenging \cite{SS}.

\begin{figure}[b] \label{Nt2}
\begin{center}
\vspace*{-0.8cm}
\includegraphics[width=6.0cm]{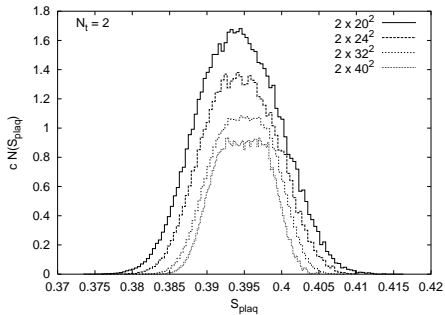}
\vspace*{-0.8cm}
\caption{Distribution of the plaquette for various lattice sizes $L^2 \times 2$.
Deviations towards a double peak characteristic of a first-order transition start to
appear at the largest sizes.}
\end{center}
\vspace*{-0.5cm}
\end{figure}

\section{NUMERICAL STUDY}

We simulate the $(2+1)d$ $SU(4)$ gauge theory using the Wilson plaquette
action on a cubic grid of size $L^2 \times N_t$, with $N_t=2,3,4$ and $L$
up to 40. To accelerate Monte Carlo evolution, we use as elementary update
a mixture of pseudo-heatbath (in all 6 $SU(2)$ subgroups) and overrelaxation in
the full $SU(4)$ group \cite{Creutz,Edinburgh}. The latter requires a similar
amount of work to 6 $SU(2)$ overrelaxation steps, but gives a larger step
size. As a result, the number of sweeps needed to decorrelate the Polyakov
loop is reduced by a factor $\sim 3$. Up to $10^6$ sweeps are performed
on each volume. 
We analyze results for various couplings $\beta$ near criticality together
with multihistogram reweighting.

This model gives a clear first-order transition on $N_t=1$ lattices.
On $N_t=2$, an old Monte Carlo study found
a second-order deconfinement transition \cite{old}. Unfortunately,
the larger sizes which we simulated revealed that
the transition is in fact first-order.
Distortions of the plaquette distribution towards a double-peak
structure appear in Fig.~1 for our largest sizes. Also, the Binder cumulant
$1 - \frac{\langle {\cal O}^4 \rangle}{3 \langle {\cal O}^2 \rangle^2}$ for
the plaquette extrapolates as $L \rightarrow \infty$ to a value below 2/3.
A likely explanation of these first-order transitions
is the vicinity of a sharp, bulk crossover at $\beta \sim 13.5$,
whereas the $N_t=1$ and $2$ transitions occur at $\beta_c \approx 8.67$ and
$14.87$ respectively.

This forced us to consider $N_t=3$ lattices, with correspondingly higher
$\beta_c$. No double-peak structure is visible on the largest
size considered ($L=32$). However, the $L \rightarrow \infty$ extrapolation
of the Binder cumulant still misses 2/3 by a small but somewhat significant 
amount. It may well be
that the transition is still weakly first-order. Our current $N_t=4$
results are consistent with a second-order transition, but do not reach 
as large volumes yet. Therefore, we present the critical exponents analyzed
from the $N_t=3$ data. 

\begin{figure}[t] 
\begin{center}
\vspace*{-0.1cm}
\hspace*{-0.2cm}
\includegraphics[width=3.8cm]{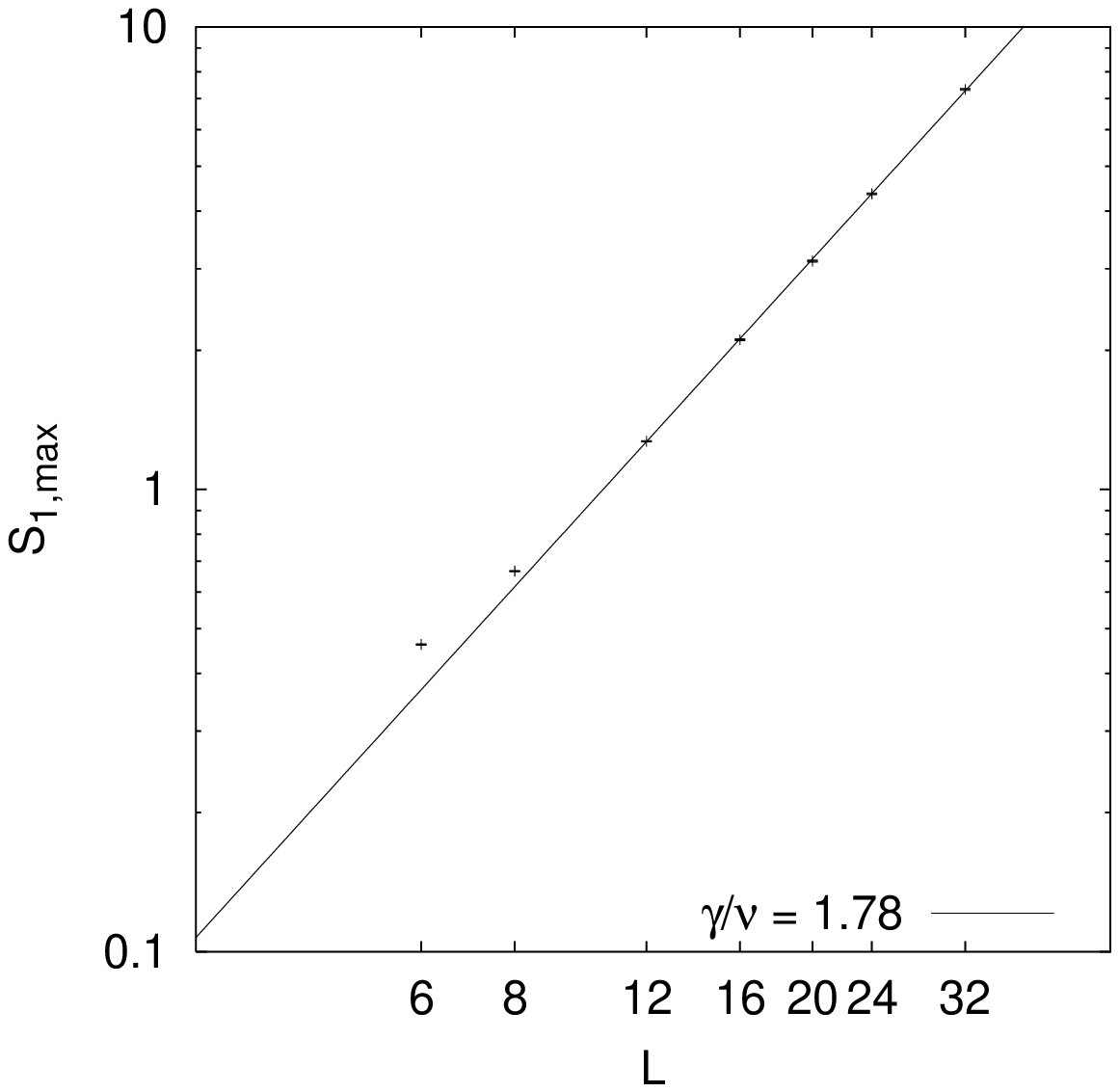}
\hspace*{-0.2cm}
\includegraphics[width=3.8cm]{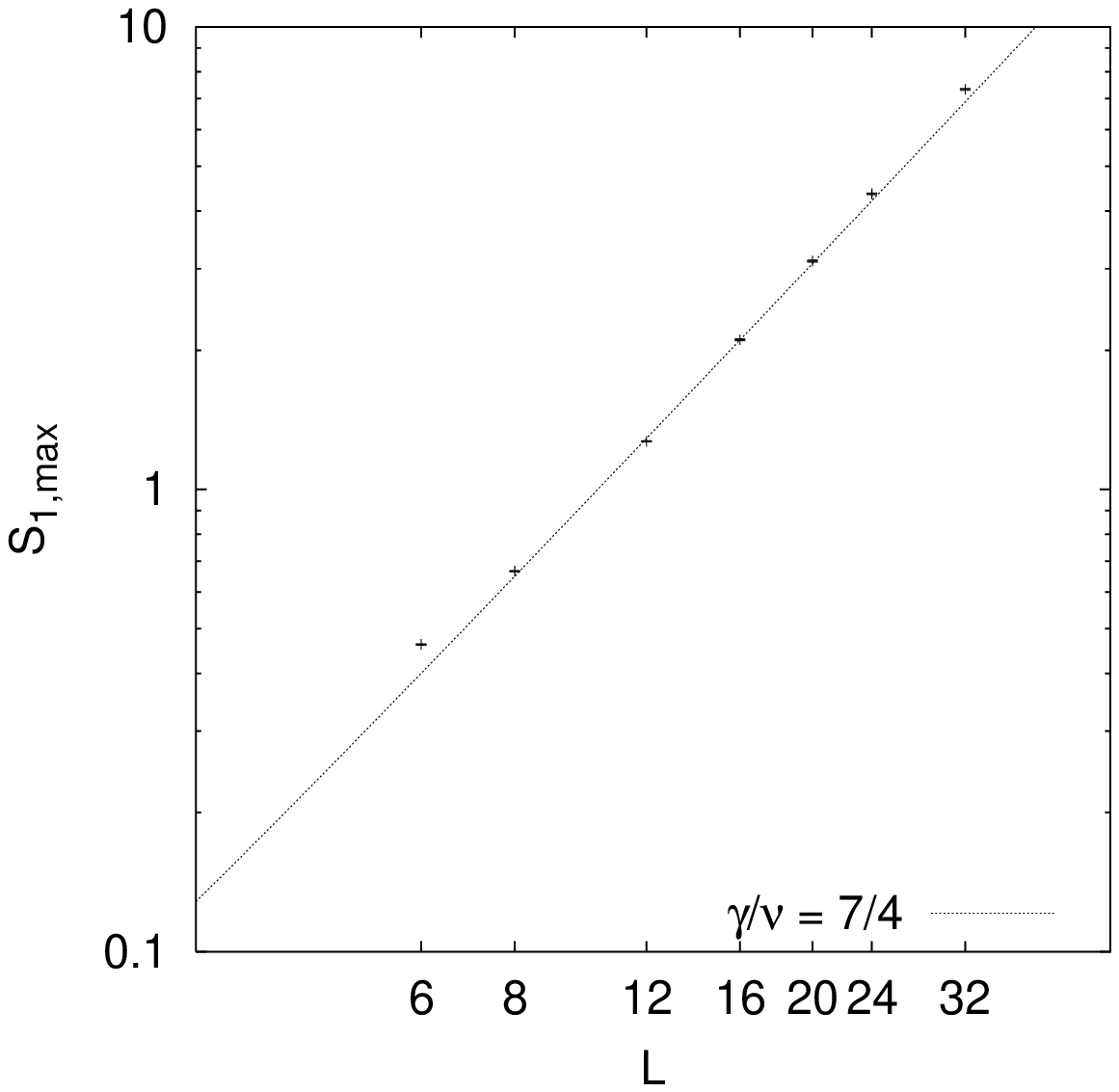}
\vspace*{-0.8cm}
\hspace*{-0.2cm}
\caption{Check of magnetic susceptibility exponent $\frac{\gamma}{\nu}=\frac{7}{4}$,
without (left) and with (right) logarithmic corrections to scaling.}
\end{center}
\vspace*{-1cm}
\end{figure}

\begin{figure}[b] 
\begin{center}
\vspace*{-0.8cm}
\hspace*{-0.2cm}
\includegraphics[width=3.8cm]{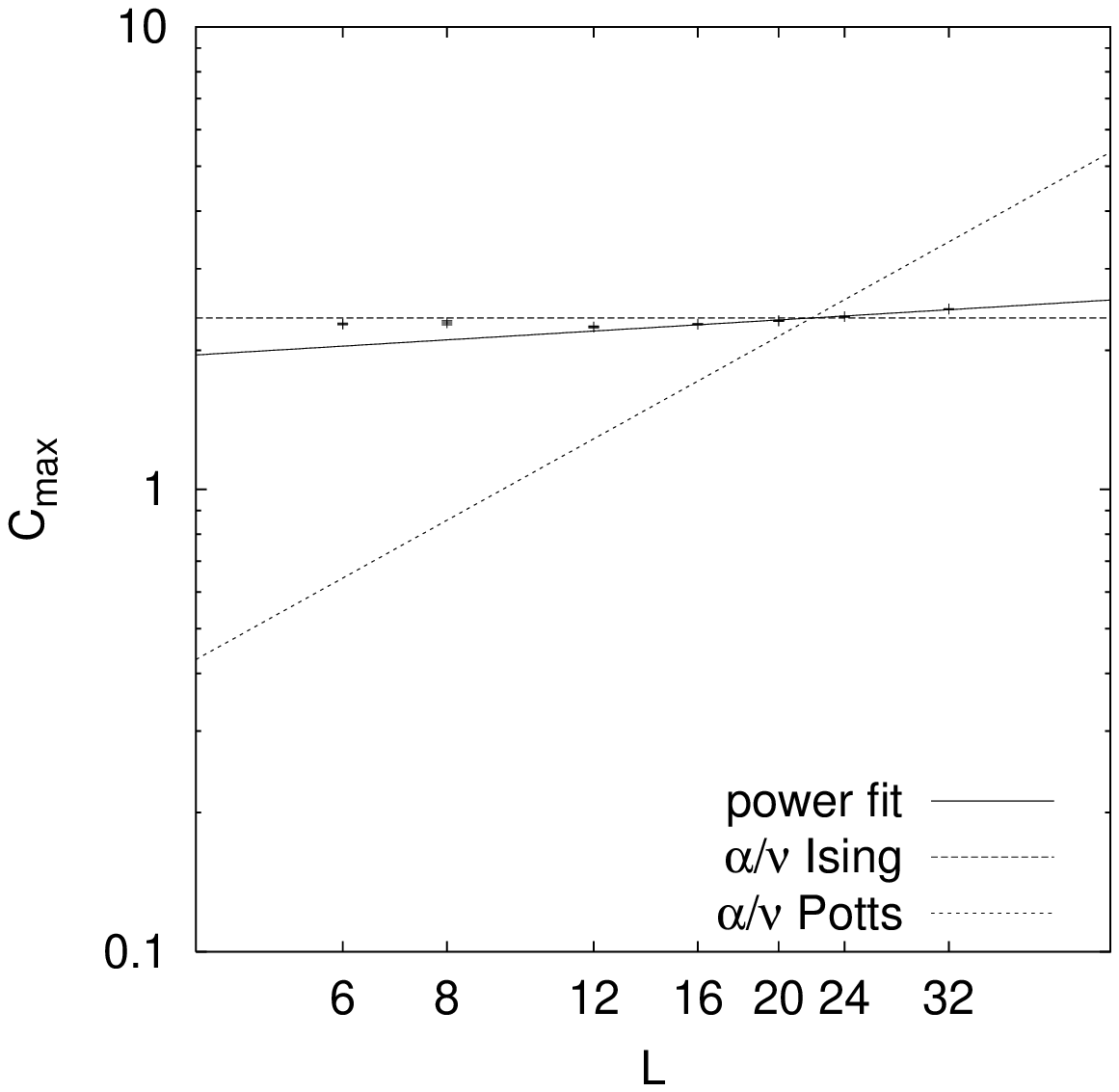}
\hspace*{-0.2cm}
\includegraphics[width=3.8cm]{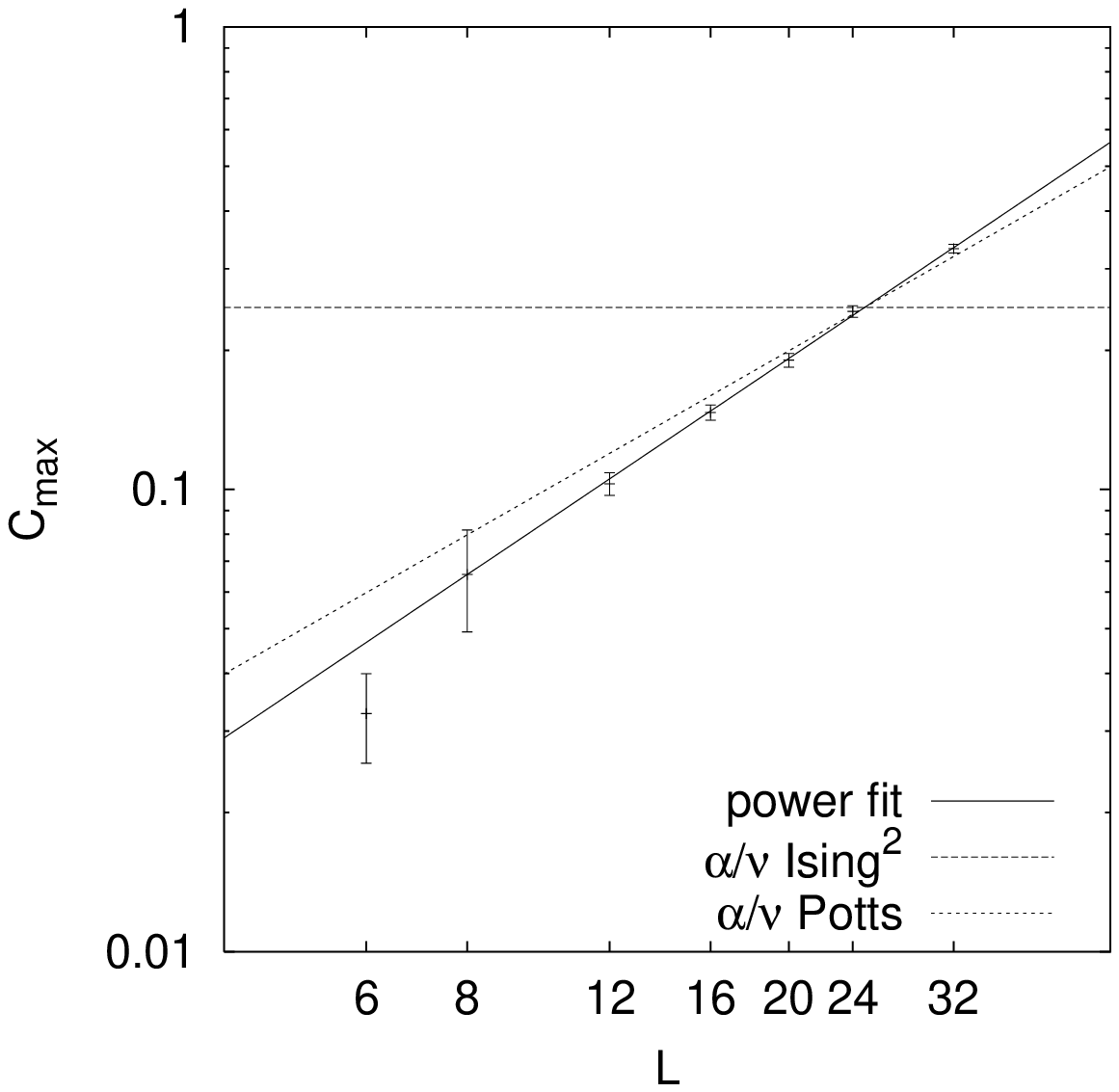}
\vspace*{-0.8cm}
\hspace*{-0.2cm}
\caption{$\frac{\alpha}{\nu}$ determined from the scaling of the specific heat, 
wihout (left) and with (right) subtraction of smooth background caused by nearby crossover.}
\end{center}
\vspace*{-0.5cm}
\end{figure}

The effect of logarithmic corrections to scaling can be seen in Fig.~2. 
They limit the usefulness of accurate, small-size data.

The bulk crossover at $\beta \sim 13.5$ has a more pernicious effect.
It dominates the behaviour of the specific heat over the singular piece,
leading to an exponent $\alpha \approx 0$ (Fig.~3 left). We subtracted the bulk
specific heat, measured on a $4^4$ lattice, to isolate the singular contribution.
The exponent then becomes consistent with the Potts case (Fig.~3 right).

All other exponents also favor the Potts case. Fig.~4 shows the scaling of $\frac{dU}{d\beta}$,
where $U$ is the Polyakov loop Binder cumulant.
$\nu$ remains consistent with the Potts value $2/3$, even allowing for a large
($> 2\sigma$) variation of $\beta_c$ by $\pm 0.01$ and logarithmic finite-size 
corrections.

Agreement among all observables is shown in Fig.~5. The pseudo-critical $\beta$'s
obtained on various volumes all extrapolate to a common thermodynamic value
$\beta_c = 20.4356(41)$, with corrections of the form 
$a L^{-\frac{1}{\nu}} (1 + b/L)$, and $\nu=2/3$, the Potts value.

\begin{figure}[t] 
\begin{center}
\includegraphics[width=6.0cm]{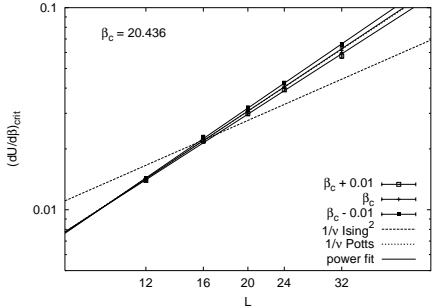} 
\vspace*{-0.8cm}
\caption{Correlation length exponent $\nu$ determined from the Polyakov loop
Binder cumulant.
The Potts value is strongly favored, even allowing for an error of $\pm 0.01$
in the determination of $\beta_c$.}
\end{center}
\vspace*{-1cm}
\end{figure}

\begin{figure}[t] 
\begin{center}
\includegraphics[width=6.0cm]{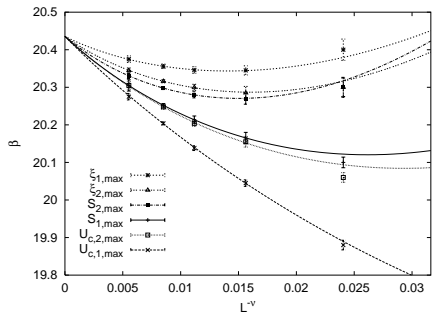}
\vspace*{-0.8cm}
\caption{Consistency of all determinations of $\beta_c$.
$\nu$ has the Potts value $2/3$; corrections to
scaling are fitted.}
\end{center}
\vspace*{-1cm}
\end{figure}

\section{CONCLUSION}

The unexpected $N_t=2$ first-order
transition, together with expected but unwelcome logarithmic finite-size 
corrections, have turned this simple problem into a numerically challenging one.

Our results need to be made more precise, by simulating larger volumes.
They also must be confirmed on a finer lattice, $N_t \ge 4$, where the transition
presumably is second-order.
In practice, we can never exclude a weak first-order transition. What we want is to 
reach physical volumes large enough to reliably determine {\em effective}
critical exponents, but small enough compared to the possibly finite
correlation length at criticality.

At this stage, the set of measured exponents favors the Potts case,
which would be perhaps the most naive guess.

Obviously, much remains to be done. The effort is
worthwhile because of the insight gained from this simple case.
Interest in Polyakov loop models has been growing, because they can supplement
dimensional reduction and provide an effective description at temperatures
nearer to $T_c$ than the latter. The difficulty is to determine their many
couplings, especially for large$-N$ $SU(N)$ theories. Lessons learnt here for
$SU(4)$ may help choose among the simplest $SU(N)$ generalizations, like
$N-$states Potts or $Z(N)$ clock models.
In that respect, our preliminary result is perhaps surprising,
because an effective $N$-state Potts model for $(3+1)d$ $SU(N)$ would
lead to $k$-string tensions independent of $k$ at $T_c$.

\end{document}